\documentstyle[preprint,aps]{revtex}
\begin{document}
\draft
\preprint{atom-ph/9509005}
\title{Squeezing in the interaction of radiation with two-level atoms}
\author{Abir Bandyopadhyay\footnote{Electronic address :
abir@iitk.ernet.in} and Jagdish Rai\footnote{Electronic address :
jrai@iitk.ernet.in}}
\address{\it Department of Physics, Indian Institute of Technology,\\
Kanpur - 208 016, INDIA.}
\date{\today}
\maketitle
\begin{abstract}
    We propose a simple experimental procedure to produce squeezing and
other non-classical properties like photon antibunching of radiation, and
amplification without population inversion. The method also decreases the
uncertainties of the angular-momentum quadratures representing the
two-level atomic system in the interaction of the two-level atoms with
quantized radiation.
\end{abstract}
    Squeezed states of quantum systems have been an active area of
interest for more than a decade \cite{walls1}. Nonclassical states of
radiation fields showing squeezing properties have been experimentally
produced in four wave mixing procedures or by passing a coherent beam
through optically nonlinear medium \cite{walls2}. It is now well
established that these states have many potential applications in
increasing the sensitivity of interferometers \cite{hillery} and in noise
free transmission of information \cite{schu}. In case of radiation,
squeezing redistributes the equal quantum noises of the quadratures
present in the minimum uncertainty or coherent states (lasers) unequally
to retain the minimum uncertainty nature. The attempt to generate squeezed
radiation is still on as a technological problem \cite{fox}. In
this letter we propose a new procedure to generate squeezing in the
interaction of a coherent radiation field with a coherent atomic beam.
Our method does not need any non-linear medium for this purpose and thus
has a much simpler theory as well as experimental requirement. The
procedure also enables to produce amplification in the radiation without
population inversion. We also show that the uncertainties of the Bloch
vector (representing atomic dipole moment) can also be reduced through the
same interaction.\\

    Schwinger developed an abstract SU(2) representation for angular
momentum systems in a Hilbert space \cite{schwinger}. Atkins and Dobson
constructed angular momentum coherent states using Schwinger
representation containing both integer and half-integer angular momentum
states in the SO(3) space \cite{ad}. We call these states as Schwinger
angular-momentum coherent (SAMC) states. These states were successfully
applied by Fonda et.al to nuclear and molecular coherent rotational states
\cite{fonda} mentioning the problem with the system of real spin-$1\over
2$ particles (fermions). However, this problem does not arise in the case
of any two-level quantum system consisted of both bosonic and fermionic
harmonic oscillator. The examples of such systems are : an admixture of
bosonic and fermionic ultracold ensemble of two-level atoms \cite{zhang}
and, two-mode radiation system \cite{yurke}. We use the first example for
the present purpose. \\

    In figure 1 we sketch the schematic experimental setup for the
generation of radiation squeezing. A hole in the mirror, deflecting the
coherent radiation (laser), injects an ultracold beam consisted of both
bosonic and fermionic atoms described by the SAMC states. The atomic
beam interacts with the radiation on the way. The Hamiltonian of the
combined atom field system in Jaynes-Cummings model \cite{jc} as (in
units of $\hbar$)
\begin{equation}
H = H_A + H_F + V_I
\end{equation}
where,$H_A = \omega_0 J_z$ is the Hamiltonian for the free two-level atom
with energy difference of $\omega_0$ between the levels, $H_F =\omega
a^\dagger a$ is the Hamiltonian of the free field with frequency $\omega$.
$V_I = g(aJ_+ + {a^\dagger}{J_-})$ is the interaction term under rotating
wave approximation (RWA) with interaction strength of $g=\sqrt{{6\pi^3
c^3 A}\over {V\omega^2}}$ \cite{butler}. $A$ is the $A$-coefficient of
the two-level atoms in the mode volume $V$. We also take $\omega =
\omega_0 $ for simplicity in the calculation.\\

	Due to the simplicity and resemblance with the classical
perception we work in the Heisenberg picture, where any operator $\hat
{O}(0)$ is transformed to $\hat{O}(t) = \exp (iHt) \hat{O} (0) \exp
(-iHt)$ as time evolves. Using Baker-Campbell-Housdroff relation for
operators $e^{\xi A} \hat{O} e^{-\xi A} = \hat{O} + \xi [A,\hat{O}] +
{{\xi^2} \over {2!}} [A,[A,\hat{O}]] + \cdots$ and retaining the terms
up to the order of $g^2$ for field operators and $g$ for angular
momentum projection operator, we were able to close the infinite series
occurring in the transformed operators. The transformed linear operators
of the field up to the order of $g^2$ are
\begin{mathletters}
\begin{eqnarray}
a(t)& = &
{e^{-i\omega t}}[a(0) - igtJ_- (0)-g^2 t^2 a(0)J_z (0)] \nonumber \\
{a^\dagger}(t)& = & e^{i\omega t} [a^\dagger(0) + igtJ_+ (0)-g^2 t^2
a^\dagger (0)J_z(0)]
\end{eqnarray}
and of the angular-momentum representing the atomic system up to the
order of $g$ are
\begin{eqnarray}
{J_+}(t) &=& e^{i\omega t} [J_+ (0) - 2igt {a^\dagger}(0){J_z}(0)]
\nonumber \\
{J_-}(t) &=& e^{-i\omega t} [J_- (0) + 2igt a(0){J_z}(0)] \nonumber \\
{J_z}(t) &=& {J_z}(0) - igta{J_z}(0)
\end{eqnarray}
\end{mathletters}
Instead of calculating the transformation of the quadratic operators we
can introduce the unity operator $\exp (-iHt)~\exp (iHt)$ between the
linear operators. Thus simply multiplying the transformed linear operators
one can find the product operators up to the order of the term in the
transformed linear operators. These transformed product operators are used
to calculate the different variances discussed below. \\

   Now, we define the initial state of the whole system in the Schrodinger
picture, i.e. the stationary state in the Heisenberg picture, as a product
of the free field and the free atomic states $\vert \Psi \rangle = \vert
\psi_F \rangle \otimes \vert \phi_A \rangle$.  Choosing both the field and
the atomic states to be coherent states as required by the experimental
proposal we can write
\begin{equation}
\vert \Psi \rangle = \vert \alpha \rangle \otimes \vert {\beta_+},
{\beta_-}\rangle
\end{equation}
where $\alpha$ is the complex parameter of the coherent radiation field
and $\beta_i$ are the complex parameters for the angular-momentum coherent
state. Calculating the matrix elements of the time evolved operators
for the above mentioned coherent-coherent states of the field and the
atomic system we found the variances of the quadratures of the field
variables up to the order of $g^2$ as
\begin{mathletters}
\begin{eqnarray}
\Delta X_1^2 (t) = {1\over 2}-&2gt\vert \alpha \vert \vert \beta_+
\vert \vert \beta_- \vert [\sin (\theta_\alpha - \theta_{\beta_+} +
\theta_{\beta_-} )&+\sin (2\omega t - \theta_{\beta_+} +
\theta_{\beta_-}) \nonumber \\
&&-2\cos (\omega t-\theta_{\alpha} )\sin (\omega t
-\theta_{\beta_+} + \theta_{\beta_-})]\nonumber \\
+& g^2 t^2 [\vert \beta_+ \vert^2 + (\vert \beta_+ \vert^2 -\vert
\beta_- \vert^2 ) \{ \vert \alpha \vert^2 \cos  2(\omega t -\theta_\alpha
)&- {\alpha \over 2}\cos (2\omega t -\theta_\alpha ) \} ]
\end{eqnarray}
\begin{eqnarray}
\Delta X_2^2 (t) = {1\over 2}-&2gt\vert \alpha \vert \vert \beta_+
\vert \vert \beta_- \vert [\sin (\theta_\alpha - \theta_{\beta_+} +
\theta_{\beta_-} )&-\sin (2\omega t - \theta_{\beta_+} +
\theta_{\beta_-}) \nonumber \\
&&+2\sin (\omega t-\theta_{\alpha} )\cos (\omega t
-\theta_{\beta_+} + \theta_{\beta_-})] \nonumber \\
+& g^2 t^2 [\vert \beta_+ \vert^2 - (\vert \beta_+ \vert^2 -\vert
\beta_- \vert^2 ) \{ \vert \alpha \vert^2 \cos  2(\omega t -\theta_\alpha
)&-{ \alpha \over 2}\cos (2\omega t -\theta_\alpha ) \} ]
\end{eqnarray}
\end{mathletters}
where $\theta$s are the phases or arguments of the complex parameters
$\alpha$ and ${\beta_\pm}$. For simplicity and a better understanding of
the results we set all the phases ($\theta$s) to be zero, which does not
reduce the importance of our results except that the linear dependencies
on $gt$ drops out. This dependency will be reported in a detailed
parametric study with all the phases. As we are interested in the
generation of non-classical properties, the second order term fulfills the
present purpose. Under this simplification the uncertainties reduce to
\begin{mathletters}
\begin{equation}
\Delta X_1^2 (t) = {1\over 2}+g^2 t^2 [\vert \beta_+ \vert^2 + (\vert
\beta_+ \vert^2 -\vert \beta_- \vert^2 )(\vert \alpha \vert^2 -{{\vert
\alpha \vert} \over 2}) \cos 2\omega t ]
\end{equation}
\begin{equation}
\Delta X_2^2 (t) = {1\over 2}+g^2 t^2 [\vert \beta_+ \vert^2 - (\vert
\beta_+ \vert^2 -\vert \beta_- \vert^2 )(\vert \alpha \vert^2 -{{\vert
\alpha \vert} \over 2}) \cos 2\omega t ]
\end{equation}
\end{mathletters}
We have plotted the time development of the uncertainties in Fig.2 for
$\nu = {\omega \over {2\pi}}= 6\times 10^{14}$Hz, $g=\omega \times 10^{-5}
$, $\vert \alpha \vert =5.0, \vert \beta_+ \vert = 1.0, \vert \beta _-
\vert =10.0$.  The uncertainties in the quadratures show oscillations due
to the sinusoidal term after a certain time and start to show squeezing
properties. The amount of squeezing become more and more over time due to
the $(gt)^2$ dependency. However, the squeezing property also oscillates
between the quadratures and at certain time intervals they return back to
the coherent state.

    We have also calculated the mean of the number of photons and the
bunching parameter $\langle {\cal B}\rangle$ (=$\langle {a^\dagger}^2 a^2
\rangle -\langle a^\dagger a\rangle^2$) for the above simplified choice of
phases to check the amplification and non-classical behavior of the
statistics of the photon number distribution. The mean number of the
photons and the bunching parameter up to the order of $g^2$
were calculated to be
\begin{equation}
\langle n(t)\rangle = \vert \alpha \vert^2 + g^2 t^2 [\vert \beta_+
\vert^2 (1+\vert \beta_- \vert^2 ) -\vert \alpha \vert^2 ( \vert \beta_+
\vert^2 -\vert \beta_- \vert^2 )]
\end{equation}
\begin{equation}
\langle {\cal B}\rangle = g^2 t^2 [\vert \beta_+ \vert^2 (3+\vert \beta_-
\vert^2 ) -(\vert \alpha \vert^2 -\vert \alpha \vert ( \vert \beta_+
\vert^2 -\vert \beta_- \vert^2 )]
\end{equation}
The expression of the mean number of photons has an extra term of the
order of $(gt)^2$ and thus if the quantity in the square bracket is chosen
to be positive, amplification of the optical signal is possible through
the interaction. The bunching parameter depends only on $(gt)^2$ with a
constant factor. If this constant factor, dependent only on the mean
number of initial photons and the mean numbers of atoms in the two states,
is chosen positive or negative, then it is possible to have
non-Poissonian (bunched for positive and antibunched for negative)
statistics of the radiation field.\\

    The angular momentum quadratures are actually the measure of the
Bloch vector or the dipole moment of the atoms. We are interested in the
position uncertainty of the Bloch vector and the population inversion in
the atomic system. So we have similarly calculated the matrix elements for
the atomic variables up to the order of $g$ but the results came out to be
very much complicated in phase dependency and can not be understood
directly. For this reason we present the results of the angular-momentum
matrix elements of present interest for the same choice of phases as in
the last section.
\begin{mathletters}
\begin{equation}
\Delta J_x^2 = {1\over 4}(\vert \beta_+ \vert^2 +\vert \beta_- \vert^2 )
+ 2gt \vert \alpha \vert \vert \beta_+ \vert \vert \beta_- \vert (\vert
\beta_+ \vert^2 - \vert \beta_- \vert^2 )\sin 2\omega t
\end{equation}
\begin{equation}
\Delta J_y^2 = {1\over 4}(\vert \beta_+ \vert^2 +\vert \beta_- \vert^2 )
- 2gt \vert \alpha \vert \vert \beta_+ \vert \vert \beta_- \vert (\vert
\beta_+ \vert^2 - \vert \beta_- \vert^2 )\sin 2\omega t
\end{equation}
\begin{equation}
\langle J_z \rangle = {1\over 2}(\vert \beta_+ \vert^2 - \vert \beta_- \vert^2)
\end{equation}
\end{mathletters}
Unlike the case of the radiation, the linear dependencies of the
quadratures of the atomic system on $gt$ does not drop out for the
simplified choice. However, the angular momentum projection or the
population difference has linear dependency on $gt$ which drops out for
the choice of phases to be zero. The correction term present there for any
phase is $+2gt\vert \alpha \vert \vert \beta_+ \vert \vert \beta_- \vert
\sin (\theta_\alpha - \theta_{\beta_+} + \theta_{\beta_-} )$. If the
phases are chosen to make this term non-zero, then population difference
can also be controlled through the interaction. This means that the
population inversion can also be controlled by the choice of the phases.
However, as we are interested in optical amplification, it is shown in the
previous subsection that the simplified choice of the phases can amplify
the optical signal even without any population inversion.\\

	We have plotted the normalized quadratures ${\cal A}_+ = {{\Delta
J_x^2}\over {\vert \langle J_z \rangle \vert}}$ and ${\cal A}_- = {{
\Delta J_y^2}\over {\vert \langle J_z \rangle \vert}}$ in the angular
momentum uncertainty relation ${{\Delta J_x^2}\over {\vert \langle J_z
\rangle \vert}}{{\Delta J_y^2}\over {\vert \langle J_z \rangle
\vert}}\geq {1\over 4}$ in fig 3 for the same choice of parameters
as in the case of field. It is noticed that the uncertainties oscillate
in time with opposite phase and come back to the initial value with a
periodicity like the field case. One more difference in the plots is
that the field starts showing the interaction effect only after some
time whereas the atoms are affected as soon the interaction starts. Also
notice that we have not started from the minimum uncertainty (though
coherent by definition) angular-momentum states where all the atoms are
coherently either in ground ($\vert \beta_+ \vert$ =0) or in excited state
($\vert \beta_- \vert$ =0) as this will drop all effects on the
quadratures of the atomic system and the it will remain at minimum
uncertainty state throughout. \\

	In connection with the experiment proposed in figure 1, the atomic
beam coming from the hole in the mirror copropagate with the coherent
radiation. The parameter $gt$ is related to the length of interaction.
The time of interaction is controlled by deflecting the atomic beam
using an atomic beam-splitter \cite{grimm} after a desired interval of
time effectively defined by the length of interaction. The atomic
beam-splitter deflects the coherent upper state atoms and the coherent
lower state atoms in two directions while the radiation passes through
without deflection. The properties of the radiation after interaction
can now be verified. Two atomic beams can be recombined at some other
place to study the properties of the atomic system.\\

    In conclusion, we have designed a simple experimental setup and
calculated the change in the field and atomic variables using the rotating
wave approximation (RWA) in Jaynes-Cummings Model for the Hamiltonian
retaining the terms of the order of square of the interaction strength for
the field variables and the terms of the order of the interaction strength
for the atomic variables. We show that under these approximations the
interaction produces squeezing in the initially coherent radiation field.
It is shown that the statistics of the photon number of the radiation
field can be made non-Poissonian through the interaction. We have also
shown amplification in the radiation field without any population
inversion in the atomic system. It is observed that the uncertainties of
the atomic system show similar oscillatory dependence over time. We have
calculated the effect of the interaction on the atomic system in coherent
angular-momentum state and prescribed a method to reduce the uncertainties
in the atomic quadrature in expense of the other.  However, it is not
possible to generate squeezed atomic systems by only increasing the time
as large time of interaction will increase the product $gt$ where the
perturbation theory breaks down.

\end{document}